\documentstyle[aps,twocolumn]{revtex}
\begin{document}
\author{P. Salgado$^{1\thanks{%
pasalgad@udec.cl}},$ M$.$ Cataldo$^{2\thanks{%
mcataldo@ubiobio.cl}}$ and S. del Campo$^{3\thanks{%
sdelcamp@ucv.cl}}$}
\address{$^1$Departamento de F\'{\i}sica, Universidad de Concepci\'{o}n,\\
Casilla 160-C, Concepci\'{o}n, Chile.\\
$^2$Departamento de F\'{\i}sica, Universidad del B\'{\i}o-B\'{\i}o,\\
Casilla 5-C, Concepci\'{o}n, Chile.\\
$^3$Instituto de F\'{i}sica, Universidad Cat\'{o}lica de Valpara\'{i}so,\\
Avda Brasil 2950, Valpara\'{i}so, Chile.}
\title{{\bf Supergravity and the Poincar\'{e} group}}
\maketitle

\begin{abstract}
An action for $\left( 3+1\right) $-dimensional supergravity genuinely
invariant under the Poincar\`e supergroup is proposed. The construction of
the action is carried out considering a bosonic lagrangian invariant under
both local Lorentz rotations and local Poincar\`e translations as well as
under diffeomorphism, and therefore the Poincar\`e algebra closes off-shell.

Since the lagrangian is invariant under the Poincar\`e supergroup, the
supersymmetry algebra closes off shell without the need of auxiliary fields.
\end{abstract}

\section{Introduction}

The construction of a supergravity theory without auxiliary fields in $%
\left( 3+1\right) $-dimensions has remained as an interesting open problem.
Recently it has been shown that the three and five dimensional
supergravities studied by Achucarro-Townsend \cite{achu} and by Chamseddine
\cite{cham} respectively, as well as the higher dimensional theory studied
by Ba\~nados-Troncoso-Zanelli \cite{ban} are Chern-Simons theories. Their
supersymmetry transformations can be written in the form
\begin{equation}
\delta _\lambda A=\nabla \lambda  \label{uno}
\end{equation}
and therefore the supersymmetry algebra closes off-shell without the need of
auxiliary fields.

It is the purpose of this paper to show that it is also possible to
construct a four dimensional supergravity without auxiliary fields provided
one chooses the bosonic lagrangian in an appropriate way. In fact, the
correct lagrangian for the bosonic sector is the Hilbert lagrangian
constructed with help of the one-form vierbein defined by Stelle \cite
{stelle} and Grignani-Nardelli \cite{grigna}. This vierbein, also called
solder form \cite{stelle}, \cite{witten} was considered as a smooth map
between the tangent space to the space-time manifold $M$ at a point $P$ with
coordinates $x^\mu ,$ and the tangent space to the internal $AdS$ space at
the point whose $AdS$ coordinates are $\zeta ^a(x)$, as the point $P$ ranges
over the whole manifold $M.$ The $fig.1$ of ref. \cite{stelle} illustrates
that such vierbein $V_\mu ^a(x)$ is the matrix of the map betweeen the
tangent space $T_x(M)$ to the space-time manifold at $x^\mu ,$ and the
tangent space $T_{\zeta (x)}\left( \left\{ G/H\right\} _x\right) $ to the
internal $AdS$ space $\left\{ G/H\right\} _x$ at the point $\zeta ^n(x),$
whose explicit form is given by the $eq.(3.19)$ of ref. \cite{stelle}$.$

Taking the limit $m\rightarrow 0$ in such $eq.(3.19)$ we obtain $V_\mu
^a(x)=D_\mu \xi ^a+e_\mu ^a$ which is the map between the tangent space $%
T_x(M)$ to the space-time manifold at $x^\mu $ and the tangent space $%
T_{\zeta (x)}\left( \left\{ ISO(3,1)/SO(3,1)\right\} _x\right) $ to the
internal Poincar\'e space $\left\{ ISO(3,1)/SO(3,1)\right\} _x$ at the point
$\zeta ^n(x)$. The same result was obtained in ref. \cite{grigna} by gauging
the action of a free particle defined in the internal Minkowski space.

\section{Gravity and the Poincar\`e group}

In this section we shall review some aspects of the torsion-free condition
in gravity. The main point of this section is to display the differences in
the invariances of the Hilbert action when different definitions of vierbein
are used.

\subsection{The torsion-free condition in general relativity}

The generators of the Poincar\`e group $P_a$ and $J_{ab\text{ }}$satisfy the
Lie algebra,
\[
\left[ P_a,P_b\right] =0;
\]
\[
\left[ J_{ab},P_c\right] =\eta _{ac}P_b-\eta _{bc}P_a;
\]
\begin{equation}
\left[ J_{ab},J_{cd}\right] =\eta _{ac}J_{bd}-\eta _{bc}J_{ad}+\eta
_{bd}J_{ac}-\eta _{ad}J_{bc}.  \label{dos}
\end{equation}
Here the operators carry Lorentz indices not related to coordinate
transformations. The Yang-Mills connection for this group is given by
\begin{equation}
A=A^AT_A=e^aP_a+\frac 12\omega ^{ab}J_{ab}.  \label{tres}
\end{equation}
Using the algebra (\ref{dos}) and the general form for the gauge
transformations on $A$
\begin{equation}
\delta A=\nabla \lambda =d\lambda +\left[ A,\lambda \right]  \label{cuatro}
\end{equation}
with
\begin{equation}
\lambda =\rho ^aP_a+\frac 12\kappa ^{ab}J_{ab},  \label{cinco}
\end{equation}
we obtain that $e^a$ and $\omega ^{ab},$ under Poincar\`e translations,
transform as
\begin{equation}
\delta e^a=D\rho ^a;\quad \delta \omega ^{ab}=0,  \label{seis}
\end{equation}
and under Lorentz rotations, as
\begin{equation}
\delta e^a=\kappa _b^ae^b;\quad \delta \omega ^{ab}=-D\kappa ^{ab},
\label{siete}
\end{equation}
where $D$ is the covariant derivative in the spin connection $\omega ^{ab}$.
The corresponding curvature is
\[
F=F^AT_A=dA+AA
\]
\begin{equation}
=T^aP_a+\frac 12R^{ab}J_{ab}  \label{ocho}
\end{equation}
where
\begin{equation}
T^a=De^a=de^a+\omega _b^ae^b  \label{nueve}
\end{equation}
is the torsion $1$-form, and
\begin{equation}
R^{ab}=d\omega ^{ab}+\omega _c^a\omega ^{cb}  \label{diez}
\end{equation}
is the curvature $2$-form

The Hilbert action
\begin{equation}
S_{EH}=\int \varepsilon _{abcd}R^{ab}e^ce^d,  \label{once}
\end{equation}
is invariant under diffeomorphism and under Lorentz rotations, but is not
invariant under Poincar\`{e} translations. In fact
\[
\delta S_{EH}=2\int \varepsilon _{abcd}R^{ab}e^c\delta e^d
\]
\begin{equation}
=2\int \varepsilon _{abcd}R^{ab}T^c\rho ^d+\text{surf. term}  \label{doce}
\end{equation}
where we see that the invariance of the action requires imposing the torsion
free condition
\begin{equation}
T^a=De^a=de^a+\omega _b^ae^b=0  \label{trece}
\end{equation}
which has effects on the algebra of local Poincar\`{e} transformations. If
we impose this condition, then the local Poincar\`{e} translations take the
form of a local change of coordinates, as we can see from the respective
transformation law
\begin{equation}
\delta _{tlp}e^a=D\rho ^a+\kappa _b^ae^b  \label{catorce}
\end{equation}
\begin{equation}
\delta _{tgc}e^a=D\rho ^a+\rho \cdot \omega _b^ae^b+\rho \cdot T^a.
\label{quince}
\end{equation}

The condition $T^a=0$ permits replacing local Poincar\`e translations by a
local change of coordinates which acts together with the local Lorentz
transformations on the gauge fields as:
\[
\delta e^a=D\rho ^a+\kappa _b^ae^b
\]
\begin{equation}
\delta \omega ^{ab}=-D\kappa ^{ab}+\varepsilon \cdot R^{ab}.
\label{dieceseis}
\end{equation}

The commutator of two local Poincar\`e translations can now be computed and
gives
\begin{equation}
\left[ \delta (\rho _2),\delta (\rho _1)\right] =\delta (\kappa )
\label{diecisiete}
\end{equation}
with $\kappa ^{ab}=\rho _1^\lambda \rho _2^\nu R_{\lambda \nu }^{ab}$.
Furthermore one finds
\begin{equation}
\left[ \delta (\kappa ^{ab}),\delta (\rho ^c)\right] =\delta (\rho ^{\prime
d})\text{ with }\rho ^{\prime a}=\rho _b\kappa ^{ba}  \label{dieciocho}
\end{equation}
and
\begin{equation}
\left[ \delta (\kappa _2),\delta (\kappa _1)\right] =\delta (\kappa _3)\text{
with }\kappa _3=\left[ \kappa _1,\kappa _2\right] .  \label{diecinueve}
\end{equation}

This means that, for non-vanishing $R^{ab}$, the local Poincar\`e
translations no longer commute, but their commutator is a local Lorentz
transformation proportional to the Riemann curvature. The rest of the
algebra is unchanged. Thus an effect of the torsion-free condition is that
the Poincar\`e algebra only closes on-shell, but does not close off-shell.

Another consequence of the torsion-free condition \cite{ban} is that it is
an equation of motion of the action, which implies that the invariance of
the action under diffeomorphisms does not result from the transformation
properties of the fields alone, but it is a property of their dynamics as
well. The problem stems from the identification between diffeomorphism,
which is a genuine invariance of the action, and local Poincar\`e
translation which is not a genuine invariance.

The torsion-free condition breaks local translation invariance in Lorentz
space, and uniquely identifies the origin of the local Lorentz frame with
the space-time point at which it is constructed.

\subsection{Gravity invariant under the Poincar\'e group}

Now we show that the formalism proposed by Stelle \cite{stelle}
Grignani-Nardelli \cite{grigna} $\left( SGN\right) $ leads to a formulation
of general relativity where Hilbert's action is invariant both under local
Poincar\`e translations and under local Lorentz transformations as well as
under diffeomorphism and therefore the Poincar\`e algebra closes off-shell.

The key ingredients of the $\left( SGN\right) $ formalism are the so called
Poincare coordinates $\xi ^a(x)$ which behave as vectors under $ISO(3,1)$
and are involved in the definition of the $1$-form vierbein $V^a$, which is
not identified with the component $e^a$ of the gauge potential, but is given
by
\begin{equation}
V^a=D\xi ^a+e^a=d\xi ^a+\omega _b^a\xi ^b+e^a  \label{un}
\end{equation}

Since $\zeta ^a,$ $e^a,$ $\omega ^{ab}$ under local Poincare translations
change as
\begin{equation}
\delta \zeta ^a=-\rho ^a,\quad \delta e^a=D\rho ^a,\quad \delta \omega
^{ab}=0;
\end{equation}
and under local Lorentz rotations change as
\begin{equation}
\delta \zeta ^a=\kappa _{\text{ }b}^a\zeta ^b,\quad \delta e^a=\kappa _{%
\text{ }b}^ae^b,\quad \delta \omega ^{ab}=-D\kappa ^{ab};
\end{equation}
we have that the vierbein $V^a$ is invariant under local Poincare
translations
\begin{equation}
\delta V^a=0
\end{equation}
and, under local Lorentz rotations, transforms as
\begin{equation}
\delta V^a=\kappa _{\text{ }b}^aV^b.
\end{equation}

The space-time metric is postulated to be
\begin{equation}
g_{\mu \nu }=\eta _{ab}V_\mu ^aV_\nu ^b
\end{equation}
with $\eta _{ab}=\left( -1,1,1,1\right) $. Thus the corresponding curvature
is given by (\ref{ocho}), but now (\ref{nueve}) does not correspond to the
space-time torsion because the vierbein is not given by $e^a$. The
space-time torsion ${\cal T}$ $^a$ is given by
\begin{equation}
{\cal T}\ ^a=DV^a=T^a+R^{ab}\xi _b
\end{equation}
The Hilbert action can be rewritten as
\begin{equation}
S_{EH}=\int \varepsilon _{abcd}V^aV^bR^{cd}
\end{equation}
which is invariant under general coordinate transformations, under local
Lorentz rotations, as well as under local Poincare translation. In fact
\begin{equation}
\delta S_{EH}=\int \varepsilon _{abcd}\delta \left( R^{ab}V^cV^d\right)
\end{equation}

\begin{equation}
\delta S_{EH}=2\int \varepsilon _{abcd}R^{ab}V^c\delta V^d=0.
\end{equation}
Thus the action is genuinely invariant under the Poincar\`e group without%
{\bf \ imposing a torsion-free condition.}

The variations of the action with respect to $\zeta ^a,$ $e^a,$ $\omega
^{ab} $ lead to the following equations:
\begin{equation}
\varepsilon _{abcd}{\cal T}\ ^bR^{cd}=0
\end{equation}

\begin{equation}
\varepsilon _{abcd}V^bR^{cd}=0
\end{equation}

\begin{equation}
\ \ \ \ \varepsilon _{[aecd}\zeta _{b]}V^eR^{cd}+\varepsilon _{abcd}V^c{\cal %
T}^d=0
\end{equation}

which reproduce the correct Einstein equations:
\[
{\cal T}\ ^a=DV^a=0
\]
\[
\varepsilon _{abcd}V^bR^{cd}=0.
\]

The commutator of two local Poincar\`e translations is given by
\begin{equation}
\left[ \delta (\rho _2),\delta (\rho _1)\right] =0
\end{equation}
i.e. the local Poincar\`e translations now commute. The rest of the algebra
is unchanged. Thus the Poincar\`e algebra closes off-shell. This fact has
deep consequences in supergravity.

\section{Supergravity in $3+1$ without auxiliary fields}

In this section we shall review some aspects of the torsion-free condition
in supergravity. The main point of this section is to show that the $\left(
SGN\right) $ formalism permits constructing a supergravity invariant under
local Lorentz rotation and under local Poincare translation as well as under
local supersymmetry transformations. This means that the superpoincare
algebra closes off shell without the need of any auxiliary fields.

\subsection{The torsion-free condition in $N=1$ supergravity}

$D=3+1$, $N=1$ supergravity is based on the superpoincar\'e algebra
\[
\left[ P_a,P_b\right] =0
\]
\[
\left[ J_{ab},P_c\right] =\eta _{ac}P_b-\eta _{bc}P_a
\]
\[
\left[ J_{ab},J_{cd}\right] =\eta _{ac}J_{bd}-\eta _{bc}J_{ad}+\eta
_{bd}J_{ac}-\eta _{ad}J_{bc}
\]
\[
\left[ J_{ab},Q^\alpha \right] =-\frac 12\left( \gamma _{ab}\right) _\beta
^\alpha Q^\beta
\]
\[
\left[ P_a,Q_\beta \right] =0
\]
\begin{equation}
\left[ Q^\alpha ,Q_\beta \right] =\frac 12\left( \gamma ^a\right) _\beta
^\alpha P_a.  \label{su1}
\end{equation}

The connection for this group is given by
\begin{equation}
A=A^AT_A=e^aP_a+\frac 12\omega ^{ab}J_{ab}+\overline{Q}\psi .  \label{su2}
\end{equation}
Using the algebra (\ref{su1}) and the general form for gauge transformations
on $A$
\begin{equation}
\delta A=D\lambda =d\lambda +\left[ A,\lambda \right]  \label{su3}
\end{equation}
with
\begin{equation}
\lambda =\rho ^aP_a+\frac 12\kappa ^{ab}J_{ab}+\overline{Q}\varepsilon
\label{su4}
\end{equation}
we obtain that $e^a$, $\omega ^{ab},$ and $\psi $ under Poincar\`e
translations, transform as
\begin{equation}
\delta e^a=D\rho ^a;\quad \delta \omega ^{ab}=0\text{; \quad }\delta \psi =0;
\label{su5}
\end{equation}
under Lorentz rotations, as
\begin{equation}
\delta e^a=\kappa _b^ae^b;\quad \delta \omega ^{ab}=-D\kappa ^{ab};\quad
\delta \psi =\frac 14\kappa ^{ab}\gamma _{ab}\psi ;  \label{su6}
\end{equation}
and under supersymmetry transformations, as
\begin{equation}
\delta e^a=\frac 12\overline{\varepsilon }\gamma ^a\psi ;\quad \delta \omega
^{ab}=0\text{; \quad }\delta \psi =D\varepsilon .  \label{su7}
\end{equation}

The consistency of the propagation of the massless Rarita-Schwinger field in
a classical gravitational background field is proved by contracting its
field equation
\begin{equation}
\gamma _5e^a\gamma _aD\psi =0  \label{su8}
\end{equation}
by the covariante derivative $D$,
\[
D\left( \gamma _5e^a\gamma _aD\psi \right) =0
\]

\begin{equation}
\gamma _5\gamma _aT^aD\psi +\gamma _5e^a\gamma _aDD\psi =0.  \label{su9}
\end{equation}

The Einstein equation and the Bianchi identity reduce (\ref{su9}) to an
identity.

Equation (\ref{su8}) does not take into account the back reaction of the
spin $3/2$ field on the gravitational field. It turns out that this back
reaction of the spin-$3/2$ field on the gravitational and on itself can be
taken into account by a generalizing of Weyl's lemma \cite{Tre}:
\begin{equation}
{\cal D}e_\mu ^a=\partial _\nu e_\mu ^a-\omega _{b\mu }^ae_\nu ^b-\frac 14%
\stackrel{\_}{\psi }_\mu \gamma ^a\psi _\nu -\Gamma _{\mu \nu }^\lambda
e_\lambda ^a=0,  \label{su10}
\end{equation}
which implies that the corresponding torsion is given by
\begin{equation}
\stackrel{\wedge }{T\text{ }}^a=T\ ^a-\frac 12\stackrel{\_}{\psi }\gamma
^a\psi .  \nonumber
\end{equation}

Supergravity is the theory of the gravitational field interacting with a
spin $3/2$ Rarita Schwinger field \cite{Des}, \cite{Freed}. In the simplest
case there is just one spin $3/2$ Majorana fermion, usually called the
gravitino. The corresponding action

\begin{equation}
S=\int \varepsilon _{abcd}e^ae^bR^{cd}+4\overline{\psi }\gamma _5e^a\gamma
_aD\psi  \label{su11}
\end{equation}
is invariant under diffeomorphism, under local Lorentz rotations and local
supersymmetry transformations, but it is not invariant under Poincar\`{e}
translations. In fact, under local Poincar\'{e} translations
\begin{equation}
\delta S=2\int \varepsilon _{abcd}R^{ab}\left( T^c-\frac 12\overline{\psi }%
\gamma ^c\psi \right) \rho ^d+\text{surf. term.}  \label{su12}
\end{equation}
\begin{equation}
\delta S=2\int \varepsilon _{abcd}R^{ab}\stackrel{\wedge }{T}^c\rho ^d+\text{%
surf}.\text{ term. }  \label{su13}
\end{equation}
The invariance of the action requires the vanishing of the torsion
\begin{equation}
\stackrel{\wedge }{T\text{ }}^a=0  \label{su14}
\end{equation}
which implies that the connection is no longer an independent variable.
Rather, its variation is given in terms of $\delta e^a$ and $\delta \psi $,
and differs from the one dictated by group theory.

An effect of the supertorsion-free condition on the local Poincar\'e
superalgebra is that all commutators on $e^a,$ $\psi $ close except the
commutator of two local supersymmetry transformations on the gravitino. For
this commutator on the vierbein one finds
\begin{equation}
\left[ \delta \left( \varepsilon _1\right) ,\delta \left( \varepsilon
_2\right) \right] e^a=\frac 12\overline{\varepsilon }_2\gamma ^aD\varepsilon
_1-\frac 12\overline{\varepsilon }_1\gamma ^aD\varepsilon _2=\frac 12D\left(
\overline{\varepsilon }_2\gamma ^a\varepsilon _1\right) .  \label{su15}
\end{equation}
with $\rho ^a=\frac 12\overline{\varepsilon }_2\gamma ^a\varepsilon _1,$ we
can write
\begin{equation}
\left[ \delta \left( \varepsilon _1\right) ,\delta \left( \varepsilon
_2\right) \right] e^a=D\rho ^a.  \label{su16}
\end{equation}
This means that, in the absence of the torsion-free condition, the
commutator of two local supersymmetry transformations on the vierbein is a
local Poincar\'e translation. However, the action is invariant by
constructi\'on under general coordinate transformations and not under local
Poincar\'e translation. The general coordinate transformation and the local
Poincar\'e translation can be identified if we impose the torsion-free
condition: since $\rho ^a=\rho ^\nu e_\nu ^a$ we can write
\[
D_\mu \rho ^a=\left( \partial _\mu \rho ^\nu \right) e_\nu ^a+\rho ^\nu
\left( \partial _\nu e_\mu ^a\right) +\frac 12\rho ^\nu \left( \stackrel{\_}{%
\psi }_\mu \gamma ^a\psi _\nu \right)
\]
\begin{equation}
+\rho ^\nu \omega _\nu ^{ab}e_{\mu b}+\rho ^\nu T_{\mu \nu }^a.
\end{equation}
This means that, if $T_{\mu \nu }^a=0,$ then the following commutator is
valid:

\[
\left[ \delta _Q\left( \varepsilon _1\right) ,\delta _Q\left( \varepsilon
_2\right) \right] =\delta _{GCT}\left( \rho ^\mu \right) +\delta
_{LLT}\left( \rho ^\mu \omega _\mu ^{ab}\right)
\]
\begin{equation}
+\delta _Q\left( \rho ^\nu \stackrel{\_}{\psi }_\nu \right)  \label{su17}
\end{equation}

where we can see that $P$ in $\left\{ Q,Q\right\} =P,$ i.e. local Poincar\'e
translation, is replaced by general coordinate transformations besides two
other gauge symmetries. The structure constants defined by this result are
field-dependent \cite{van}, which is a property of supergravity not present
in Yang-Mills Theory .

The commutator of two local supersymmetry transformations on the gravitino
is given by
\[
\left[ \delta \left( \varepsilon _1\right) ,\delta \left( \varepsilon
_2\right) \right] \psi =\frac 12\left( \sigma _{ab}\varepsilon _2\right)
\left[ \delta \left( \varepsilon _1\right) \omega ^{ab}\right]
\]
\begin{equation}
-\frac 12\left( \sigma _{ab}\varepsilon _1\right) \left[ \delta \left(
\varepsilon _2\right) \omega ^{ab}\right] .  \label{su18}
\end{equation}

The condition $\stackrel{\wedge }{T}^a=0$ leads to $\omega ^{ab}=\omega
^{ab}(e,\psi )$ which implies that the connection is no longer an
independent variable and its variation $\delta \left( \varepsilon \right)
\omega ^{ab}$ is given in terms of $\delta \left( \varepsilon \right) e^a$
and $\delta \left( \varepsilon \right) \psi .$ Introducing $\delta \left(
\varepsilon \right) \omega ^{ab}(e,\psi )$ into (\ref{su18}) we see that,
without the auxiliary fields, the gauge algebra does not close, as shows the
eq. $\left( 10\right) $ of ref.\cite{van}. Therefore the condition $%
\stackrel{\wedge }{T}^a=0$ not only breaks local Poincare invariance, but
also the supersymmetry transformations. We show that, if we use the $\left(
SGN\right) $ formalism, the gauge algebra closes without the auxiliary
fields because it is not necessary to impose the torsion free condition.

\subsection{Supergravity invariant under the Poincar\'e group}

Analogous to the pure gravity case, the action for supergravity in $3+1$
dimensions is not invariant under local Poincar\'e translations. The
invariance of the action requires, in accord with 1.5 formalism, the
vanishing of the torsion $\stackrel{\wedge }{T\text{ }}^a,$ which implies
that the connection is no longer an independent variable. Rather, its
variation is given in terms of $\delta e^a$ and $\delta \psi $, and differs
from the one dictated by group theory. As a consequence the supersymmetry
algebra, acting on the spinor field, closes off-shell only with auxiliary
fields. How to construct supergravity in four dimensions without auxiliary
fields is an interesting open problem. Now we will show that this
construction is possible.

The massless Rarita-Schwinger field is a spin $3/2$ field that can be
described by a Majorana vector-spinor $\psi _\mu $ that satisfies the
equation
\begin{equation}
\gamma _5V^a\gamma _ad\psi =0
\end{equation}
where now,
\begin{equation}
V^a=D\xi ^a+e_\mu ^a.
\end{equation}

The coupling of this field to a gravitational field satisfying the free
Einstein equations is achieved by minimal coupling. According to this
prescription, it is possible to generalize the free spin-$3/2$ equation
consistently to include interaction with a gravitational background field,
as
\begin{equation}
\gamma _5V^a\gamma _aD\psi =0.  \label{r1}
\end{equation}

The consistency of the propagation of a spin $3/2$ particle in a classical
gravitational background field is proved by contracting (\ref{r1}) with
another derivative $D$:
\[
D\left( \gamma _5V^a\gamma _aD\psi \right) =0
\]

\begin{equation}
\gamma _5\gamma _a{\cal T}\text{ }^aD\psi +\gamma _5V^a\gamma _aDD\psi =0.
\label{s14}
\end{equation}
The Einstein equation and the Bianchi identity reduce (\ref{s14}) to an
identity. Of course, using the free Einstein equation implies that we have
not taken into account the back reaction of the spin $3/2$ field on the
gravitational field. In that sense the gravitational field here is just a
fixed classical background field. The more general situation, in which both
the gravitational field and the Rarita-Schwinger field are dynamical, is the
situation encountered in supergravity.

Equation (\ref{r1}) does not take into account the back reaction of the spin
$3/2$ field on the gravitational field; namely, the spin-$3/2$ field itself
can act as a source for the gravitational field. Since the spin $3/2$ field
is coupled to the gravitational field through the covariant derivative $D$,
this also induces a coupling of the gravitino field with itself. It turns
out that this back reaction of the spin-$3/2$ field on the gravitational and
on itself can be taken into account by generalizing Weyl's lemma \cite{Tre}:
\begin{equation}
{\cal D}_\nu V_\mu ^a=\partial _\nu V_\mu ^a-\omega _{b\mu }^aV_\nu ^b-\frac
14\stackrel{\_}{\psi }_\mu \gamma ^a\psi _\nu -\Gamma _{\mu \nu }^\lambda
V_\lambda ^a=0.  \label{s15}
\end{equation}
This implies that the corresponding torsion is given by
\begin{equation}
\stackrel{\wedge }{{\cal T}\text{ }}^a={\cal T}\ ^a-\frac 12\stackrel{\_}{%
\psi }\gamma ^a\psi .  \nonumber
\end{equation}

Within the$\left( SGN\right) $ formalism the action for supergravity can be
rewritten as

\begin{equation}
S=\int \varepsilon _{abcd}V^aV^bR^{cd}+4\overline{\psi }\gamma _5V^a\gamma
_aD\psi  \label{s17}
\end{equation}
which is invariant under local Lorentz rotations
\begin{equation}
\delta V^a=\kappa _{\text{ }b}^aV^b\text{ ; \qquad }\delta \psi =\frac 14%
\kappa ^{ab}\gamma _{ab}\psi ;  \label{s21}
\end{equation}
i.e. under $\delta \omega ^{ab}=-D\kappa ^{ab}$; $\delta e^a=\kappa _{\text{
}b}^ae^b$; $\delta \psi =\frac 14\kappa ^{ab}\gamma _{ab}\psi $; $\delta \xi
^a=\kappa _{\text{ }b}^a\xi ^b$; under local Poincar\'e translations
\begin{equation}
\delta V^a=0\text{ ; \qquad }\delta \psi =0;  \label{s22}
\end{equation}
i.e. under $\delta \omega ^{ab}=0$ ; $\delta e^a=D\rho ^a;$ $\delta \psi =0$
; $\delta \xi ^a=-\rho ^a$; and under local supersymmetry transformations
\begin{equation}
\delta V^a=i\overline{\varepsilon }\gamma ^a\psi \text{ ; \qquad }\delta
\psi =D\varepsilon ;\qquad  \label{s23}
\end{equation}
i.e. under $\delta \omega ^{ab}=0$; $\delta e^a=i\overline{\varepsilon }%
\gamma ^a\psi ;$ $\delta \psi =D\varepsilon ;$ $\delta \xi ^a=0.$ \qquad

This means that the action (\ref{s17}) is invariant without the need to
impose a torsion-free condition. We can see that, in the local Poincar\`e
superalgebra, all commutators on $e_{\text{ }}^a$,$\psi $ close including
the commutators of two local supersymmetry transformations on the vierbein
and on the gravitino. In fact, for this commutator on the vierbein one
finds:
\begin{equation}
\left[ \delta \left( \varepsilon _1\right) ,\delta \left( \varepsilon
_2\right) \right] e^a=D\rho ^a  \label{s24}
\end{equation}
where $\rho ^a=\overline{\varepsilon }_2\gamma ^a\varepsilon _1,$ i.e. the
commutator of two local supersymmetry transformations on the vierbein is a
local Poincar\'e translation.

The commutator of two local supersymmetry transformations on the gravitino
is given by
\[
\left[ \delta \left( \varepsilon _1\right) ,\delta \left( \varepsilon
_2\right) \right] \psi =\frac 12\left( \sigma _{ab}\varepsilon _2\right)
\left[ \delta \left( \varepsilon _1\right) \omega ^{ab}\right]
\]
\begin{equation}
-\frac 12\left( \sigma _{ab}\varepsilon _1\right) \left[ \delta \left(
\varepsilon _2\right) \omega ^{ab}\right] =0  \label{s25}
\end{equation}
because now the connection is an independent variable i.e. $\delta \left(
\varepsilon \right) \omega ^{ab}=0$ in accord with the group theory. This
proves that, if we use the $\left( SGN\right) $ formalism$,$ the gauge
algebra closes without the auxiliary fields, because it is not necessary to
impose the torsion-free condition.

In the context of a genuinely first order formalism, i.e. where the spin
connection $\omega _{\text{ }b\mu }^a$ transforms independently of the
graviton field $e_{\text{ }\mu }^a$, of the gravitino field $\psi ,$ and of
the Poincare field $\xi ,$ the field equations can be obtained by varying (%
\ref{s17}) with respect to $\overline{\psi }_\mu $ , $e_\mu ^a$ and with
respect to $\omega _\mu ^{ab}$

\begin{equation}
\gamma _5V^a\gamma _aD\psi =0,  \label{s18}
\end{equation}
$\qquad $

\begin{equation}
\varepsilon _{abcd}V^bR^{cd}+2\overline{\psi }\gamma _5\gamma _aD\psi =0,
\label{s19}
\end{equation}

\begin{equation}
\varepsilon _{abcd}V^c\stackrel{\wedge }{{\cal T}\text{ }}^d+\varepsilon
_{[aecd}\xi _{b]}V^eR^{cd}+2\xi _b\overline{\psi }\gamma _5\gamma _aD\psi =0.
\label{s20}
\end{equation}

\section{Comments}

The off-shell closure of the gauge algebra in $\left( 3+1\right) $%
-dimensional supergravity is a problem that was studied a long time ago by
Kaku, Townsend, van Nieuwenhuizen \cite{kaku}, \cite{van}. They found that
the action for conformal supergravity is invariant under local $K$-gauges if
the $P$-curvature vanishes $T_{\mu \nu }^a=0$ and if the $Q$-curvature is
chiral dual $R_{\mu \nu }(Q)+\frac 12\varepsilon _{\mu \nu }^{\rho \sigma
}R_{\rho \sigma }(Q)\gamma _5=0$. Local $Q$-invariance follows if one
imposes one more constraint $R_{\mu \nu }(Q)\sigma ^{\mu \nu }=0.$ The
torsion free condition leads to $\omega _\mu ^{ab}=\omega _\mu ^{ab}(e,\psi
) $ and the duality constraint leads to $\phi _\mu =\phi _\mu (K).$ The
transformation of $\omega _\mu ^{ab}$ and $\phi _\mu $ obtained by
application of the chain rule differs from the group transformation. Using
the cyclic identity $\varepsilon ^{\mu \nu \rho \sigma }\gamma _\nu
\overline{R}_{\rho \sigma }(Q)=0$ one obtains the transformation of $\omega
_\mu ^{ab}$: $\delta ^{\prime }\omega _\mu ^{ab}=\frac 12\left[
R^{ab}(Q)\gamma _\mu \varepsilon \right] .$ This transformation permits
showing that in conformal supergravity the gauge algebra closes without the
need to use the field equations.

In conformal supergravity, as in usual supergravity, the Poincar\`e
translation is not a symmetry of the action, but one must consider instead
general coordinate transformations. The reason why the Poincar\`e algebra
does not close on $\psi $ is that the cyclic identity $\varepsilon ^{\mu \nu
\rho \sigma }\gamma _\nu \overline{R}_{\rho \sigma }(Q)=0$ is not available
to cast $\delta \omega _\mu ^{ab}$ into the simple form $\delta ^{\prime
}\omega _\mu ^{ab}$ that appears in conformal supergravity.

We have shown in this work that the successful formalism used by Stelle \cite
{stelle} and Grignani-Nardelli \cite{grigna} to construct an action for $%
\left( 3+1\right) $-dimensional gravity invariant under the Poincar\`e group
can be generalized to supergravity in $\left( 3+1\right) $-dimensions. The
extension to other even dimensions remains an open problem. The main result
of this paper is that we have shown that the $\left( SGN\right) $ formalism
permits constructing a supergravity invariant under local Lorentz rotation
and under local Poincare translation as well as under local supersymmetry
transformations, which means that the gauge algebra closes off shell without
the need of any auxiliary fields.

The action reduces to the usual action for supergravity if we choice $\xi
^a=0,$ i.e. if we identify uniquely the origin of the local Lorentz frame
with the space-time point at which it is constructed.

{\bf Acknowlegments}

This work was supported in part by FONDECYT through Grants \# 1010485 and \#
1000305, in part by Direcci\'on de Investigaci\'on, Universidad de
Concepci\'on through Grant \# 98.011.023-1.0, and in part by UCV through
Grant UCV-DGIP \# 123.752/00.

\end{document}